# Lasing of ambient air with microjoule pulse energy pumped by a multi terawatt IR femtosecond laser

Guillaume Point, Yi Liu*, Yohann Brelet, Sergey Mitryukovskiy, Pengji Ding, Aurélien Houard, and André Mysyrowicz[+]

Laboratoire d'Optique Appliquée, ENSTA ParisTech/CNRS/Ecole Polytechnique, 828, Boulevard des Maréchaux, Palaiseau, 91762, France
*Corresponding author: yi.liu@ensta-paristech.fr; [+]andre.mysyrowicz@ensta-paristech.fr

We report on the lasing action of atmospheric air pumped by an 800 nm femtosecond laser pulse with peak power up to 4 TW. Lasing emission at 428 nm increases rapidly over a small range of pump laser power, followed by saturation above ~ 1.5 TW. The maximum lasing pulse energy is measured to be 2.6 μJ corresponding to an emission power in the MW range, while a maximum conversion efficiency of $3.5 \times 10^{-5}$ is measured at moderate pump pulse energy. The optical gain inside the filament plasma is estimated to be excess of 0.7/cm. The lasing emission shows a doughnut profile, reflecting the spatial distribution of the pump-generated white-light continuum that acts as a seed for the lasing. We attribute the pronounced saturation to the defocusing of the seed in the plasma amplifying region and to the saturation of the seed intensity.

Stimulated emission from air plasma columns has attracted wide attention because of its potential application for remote sensing [1-12]. In 2003, Q. Luo et al. reported a weak backward Amplified Spontaneous Emission (ASE) from filaments formed in air [1]. Using mid-infrared lasers at 3.9 μm or 1.03 μm and a mixture of high pressure nitrogen and argon gases, D. Kartashov et al. established in 2011 the occurrence of an intense backward stimulated emission at 337 and 357 nm [2]. These emission lines correspond to transitions between the third excited state $C^3\Pi_u^+$ and second excited state $B^3\Pi_g^+$ of neutral $N_2$ molecules, similar to the traditional nitrogen laser [13]. In the mean time, backward lasing in air due to optical dissociation of $O_2$ molecules and subsequent two photon excitation of oxygen atoms with a UV laser pulse has also been reported [3].

During the same period, J. Yao and co-workers reported several strong stimulated emission lines in the forward direction in ambient air or pure nitrogen [7]. These radiation lines at 391, 428, and 471 nm can be activated either by a wavelength tunable mid-infrared laser (1.29 μm - 4 μm) or an 800 nm femtosecond pulse together with an additional seed pulse at a proper wavelength [7-9]. Forward single pass lasing of these emission lines has been attributed to the following mechanism: rapid population inversion between the $B^2\Sigma_u^+$ and $X^2\Sigma_g^+$ states of the $N_2^+$ induced by the pump laser, followed by seeded amplification. The seed is either the 3rd and/or 5th harmonic signal of the mid-infrared pump or an external seed pulse around 400 nm in the case of the 800 nm pump.

Very recently, three groups have independently reported self-seeded forward lasing action from $N_2^+$ in air or pure nitrogen pumped by just one single pulse at 800 nm [10-12]. Seeding in this case is provided by the generation of 2nd harmonic and supercontinuum white-light inside the gas [10-12]. Discovery of the self-seeded air lasing effect pumped by the widely available 800 nm femtosecond laser simplifies the experimental setup and offers the possibility of scaling up to Terawatt (TW) class pump lasers. In previous experiments so far, however, the IR pump power was limited to a few tens of GW [7-12]. Consequently, the energy of the lasing emission was too weak to be measured.

In this letter, we study self-seeded lasing from laser filaments in ambient air driven by 800 nm femtosecond pulses with a peak power of 4 TW. In the experiment, we used the Enstamobile laser system which delivers 50 fs laser pulses at a repetition rate of 10 Hz with pulse energy up to 300 mJ. The beam exhibits a super-Gaussian profile with a diameter of 40 mm (FMHM). The output pulses were focused by a convex lens of focal length $f$ = 100, 150 and 200 cm. Best results were obtained with the $f$ = 100 cm lens. After the plasma, two pieces of glass filters (BG 40) were installed in the beam to remove the strong infrared driving pulse and the accompanying white light with wavelength longer than 650 nm. For detection, a convex lens of $f$ = 10 cm was used to collect the radiation into a fibre spectrometer (HR 4000, Ocean Optics). The energy of the lasing emission was measured with a sensitive laser pulse energy meter (model: OPHIR, NOVA, PE9-C). We also measured the spatial profile of the lasing radiation with a charge-coupled device (CCD) camera.

In Fig. 1, we present the measured spectrum of the forward UV radiation as a function of incident laser pulse energy. At ambient pressure, the only line showing significant gain is at 428 nm. It appears as a narrow-band emission peak superimposed upon the tail of a broad continuum extending to larger wavelength for pump pulse powers larger than 150 GW (7.5 mJ). Upon increase of the incident laser power, the radiation at 428 nm increases progressively and shows saturation beyond $P$ = 1.5

TW, as shown in Fig. 2. The measured width of this radiation was 1.2 nm, limited by the resolution of the spectrometer. Note that the real spectral width is on the order of ~ 0.1 nm [14]. This radiation line has been identified as due to the transition between the second excited state ($B^2\Sigma_u^+(v=0)$) of the $N_2^+$ and its ground state ($X^2\Sigma_g^+(v'=1)$), where $v$ and $v'$ denote the molecular vibrational quantum number of the corresponding states [10-12]. At the same time, the broad-band white light emission was also observed to increase gradually with a noticeable blue shift of its edge. In all our experiment, lasing radiation at 391 nm was rarely observed and was very weak compared to that around 428 nm.

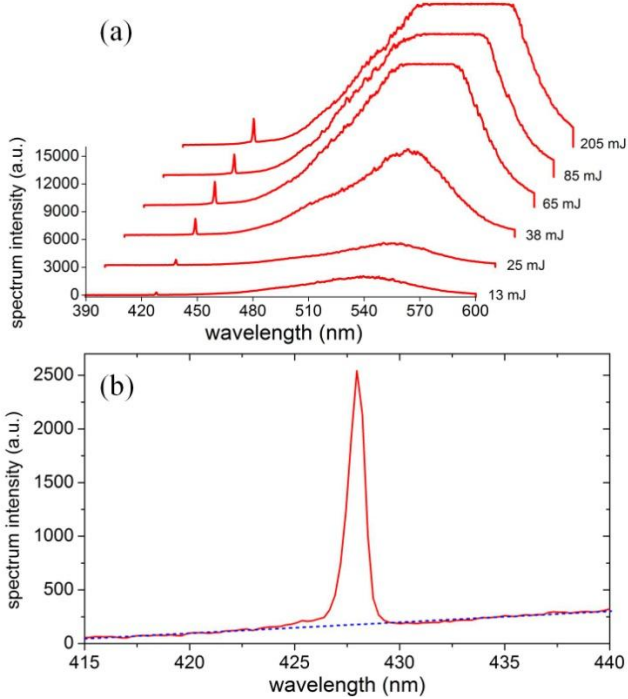

Fig. 1. (a), Spectrum of the forward emission after filamentation of laser pulses in ambient air. The focal length was 100 cm. (b), Zoomed spectrum around 428nm in the case of $E_{in}$ = 65 mJ. The dotted line shows the estimated contribution of the supercontinuum white-light.

We estimated the optical gain inside the filaments plasma as $g = \ln[I_{out}/I_{seed}]/L$, where $I_{out}$, $I_{seed}$, and $L$ are the output lasing intensity, the seed lasing intensity and the effective length of the plasma amplifier. In Fig. 1(b), we present the zoomed forward spectrum in the case of $E_{in}$ = 65 mJ. The optical gain is therefore estimated to be $g = \ln[2546/172]/3.95 = 0.7/cm$. This gain value is probably seriously underestimated here due to the limited spectral resolution of the spectrometer, which results in an order of magnitude decrease of the observed peak emission intensity at 428 nm.

We measured the energy of the lasing radiation at 428nm with the laser energy meter equipped with an interference filter at 428 nm (10 nm bandwidth). The net energy of the lasing radiation was calibrated with careful consideration of the spectrum weight of the lasing emission in the total transmitted radiation, the reflection loss of the exit fused silica windows and the collecting lens, the transmission property of the BG 40 color filter and of the interference filter. A calibrated energy of 2.6 µJ was deduced for the 428 nm lasing emission with $P_{in}$ = 2.4 TW. The maximum conversion efficiency was found to be $3.5 \times 10^{-5}$ for a pump pulse power of 1.3 TW.

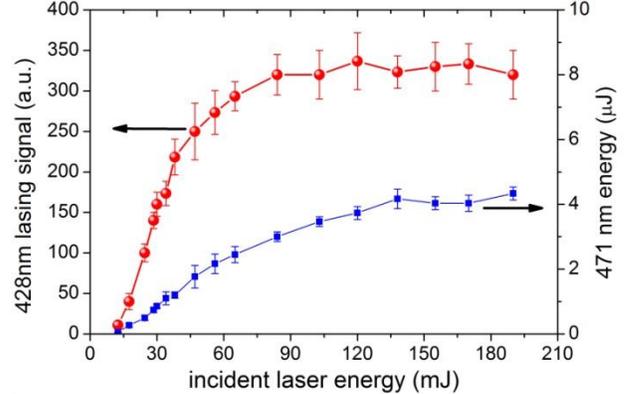

Fig.2. Dependence of the 428 nm lasing signal and the 471nm white-light spectrum component energy as a function of the incident laser pulse energy.

We measured the spatial profile of the 428 nm lasing radiation with a CCD camera equipped with the same interference filter. The results are presented in Fig. 3(a)-3(d). The radiation shows a doughnut-shaped pattern, which becomes increasingly irregular with higher pump power due to the appearance of localized intense spots. The opening angle shows a slight increase for higher incident pulse energies. In Fig. 3(a), the divergence angle of the emission was estimated to be 18.6 mrad. Such a small angle confirmed again the stimulated radiation nature of this emission. In a previous work, we tentatively attributed this doughnut-shaped radiation to the conical shape of the seeding white-light pulses [10]. Here we examine this hypothesis in more details. First, we monitored the spatial distribution of the broadband continuum at a wavelength close to the seed at 428 nm. The profile of the white-light component around 471 nm (10 nm bandwidth), shown in Fig. 3(e), displays a similar doughnut shape. Next, we checked the profile of the entire white-light emission on a screen installed 3 meters away from the geometric focus of the $f$ = 1 m focusing lens. The result presented in Fig. 3(f), shows a darker central area surrounded by visible radiation with different colors. Based on the above two observations, we conclude that the spatial distribution of the white-light seeding pulse is at the origin of the doughnut shaped 428 nm single pass lasing. Such conical emission has been widely observed during filamentation in air and solid and several mechanisms have been proposed for its explanation [15].

As mentioned above, the current understanding of the forward lasing in air is the amplification of the white-light spectral component at 428 nm in the presence of population inversion between the two related states of $N_2^+$ in the filaments plasma [7-12]. Four key factors for the single pass lasing from filaments can be identified, namely the incident seed pulse energy, the density of population inversion, the volume of the plasma amplifier, and the spatial overlap between the seeding pulse and the amplifying plasma media. The role of the first three factors are expected to be similar to that in a traditional laser amplifier [16], while the last factor is particular for this filaments amplifier due to the non uniform transverse profile of the laser intensity [17], the plasma nature of the amplifier, and the formation of multiple filaments. In the following, we examine the evolution of these factors as a function of the incident IR pulse energy.

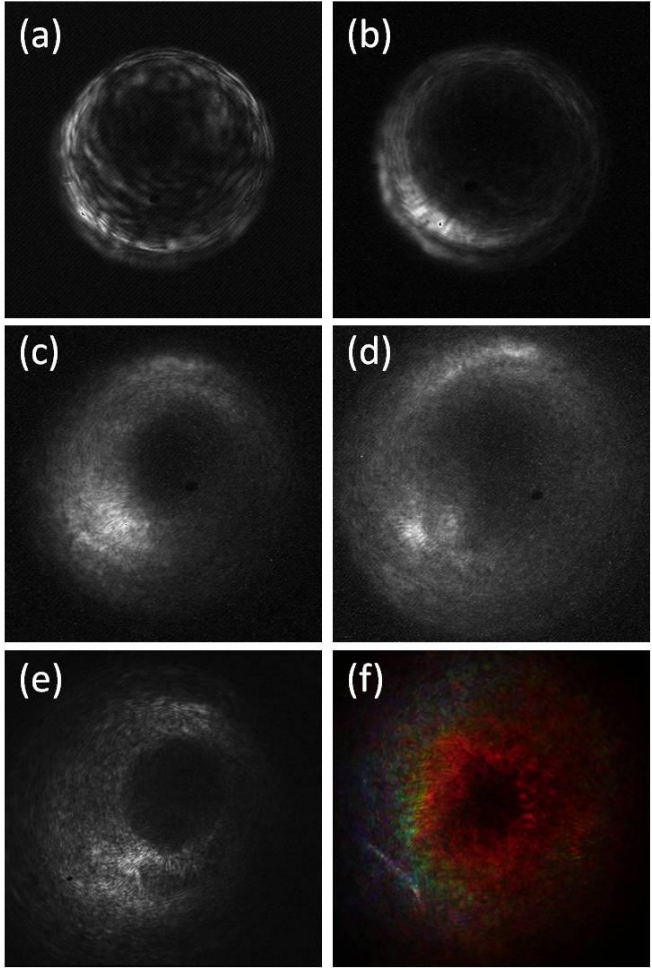

Fig. 3. (a)-(d), Spatial profiles of the 428 nm lasing radiation for different incident laser energies. The incident laser pulse energy was 18, 30, 103, and 205 mJ, respectively. (e), Profile of the white-light component around 471 nm. (f), Image of the total white-light conical emission on a screen situated 3 m away from the focal length. For (e) and (f), the incident laser energy was 80 mJ.

We first checked the volume of the filaments plasma amplifier. The images of the filaments taken by a CCD camera are presented in Fig. 4 as a function of incident laser pulse energy. The length and width of the filaments are seen to increase at higher pulse energy. When the pulse energy exceeds 84 mJ, clear evidence for multiple filaments formation at the initial stage of filamentation is observed (Fig. 4 (e)-4(g)). We numerically integrated each rows (lines) of the fluorescence plasma track and defined the plasma channel length (width) by $1/e^2$ of the maximum value. The results are presented in Fig. 5 (a). The plasma volume is estimated as $V = \pi(d/2)^2 \times L$, where $d$ and $L$ are the length and width of the plasma respectively. We show the results in Fig. 5(b), where a superlinear increase is observed at higher pump pulse energies.

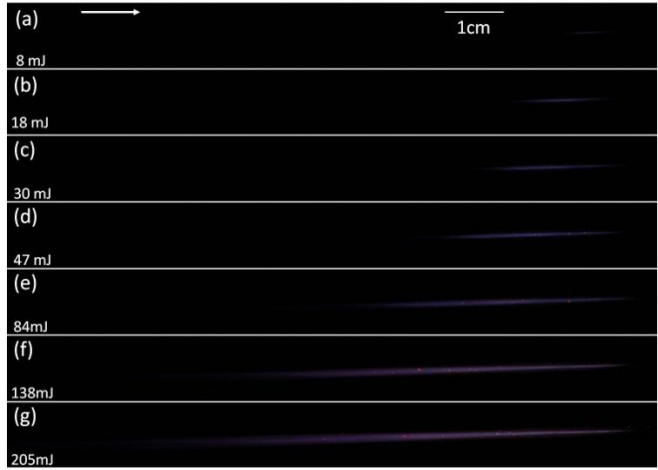

Fig. 4. Images of the filaments with different incident pulse energies, as indicated in each panel. The arrow in (a) denotes the direction of laser propagation.

We next discuss the density of population inversion. The mechanism for population inversion is still a matter of controversy [7-12]. As a result, a direct evaluation of population inversion inside the plasma channel is impossible at this stage. Nevertheless, it is reasonable to assume that the population inversion is related to the pump laser intensity, based on the observation that the population inversion is achieved within the driven pulse duration [8, 9]. It is known that the laser intensity inside a single filament remains almost unchanged, or "clamped" for a given focusing geometry above a critical power threshold, due to the interplay between the Kerr self-focusing effect and the plasma defocusing effect [18]. In the case of many filaments compressed to a bundle, the peak intensity can increase above the clamping value, as we will report elsewhere. Therefore, we expect that the density of population inversion is either roughly constant or even increases in our experiment. Since the amplifying medium increases while its population inversion density increases or stays constant, the saturation must have another origin.

We now discuss the dependence of the seed pulse energy with pump energy. Lasing action occurs at the spectrum position of the seed pulse, which prohibits a net measurement of the seed pulse energy as a function of pump laser energy in the absence of gain. Therefore, we analyzed the evolution of the seed pulse energy at another closely spaced wavelength, at 471 nm with a bandwidth of 10 nm.

The results are presented in Fig. 2. Upon increase of the pump pulse energy, the white light spectrum around 471 nm increases and progressively saturates, but the saturation is less pronounced than for the lasing signal. Therefore, the strong saturation of the 428 nm lasing radiation for $P_{in}$ > 1.5 TW is not due solely to the saturation of the seeding pulse.

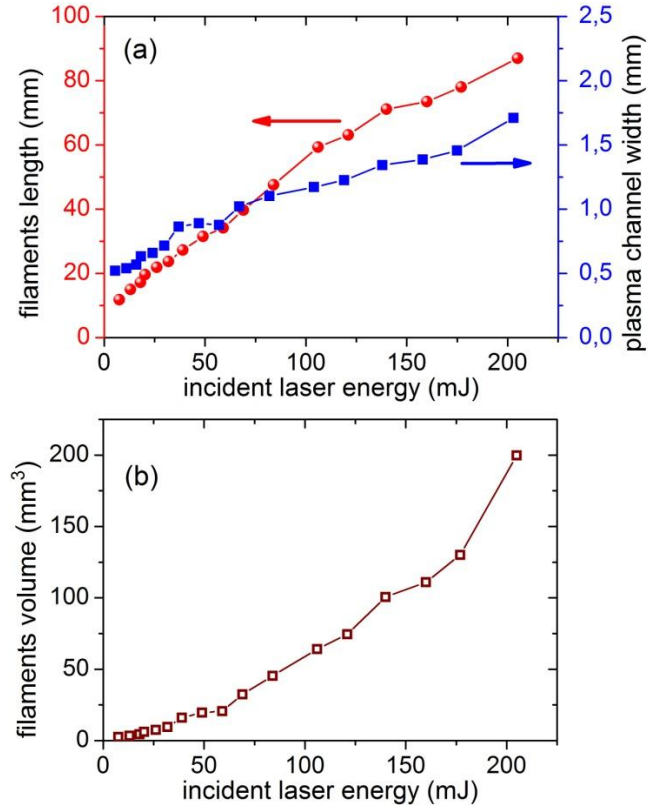

Fig. 5. (a), Length and width of the plasma channel as a function of the incident laser energy. (b), Calculated filaments volume.

Finally, we discuss the degree of spatial overlap between the seeding pulse and active amplifying region. Since the amplifying region is a plasma, it acts as a defocusing medium. Under the joint action of self-focusing and plasma defocusing, the fundamental laser pulse and the accompanying white-light supercontinuum can form complex transverse distribution inside the filaments [17], which deteriorates the spatial overlap between the seeding pulse and amplifying plasma. As presented in Fig. 3(e) and 3(f), the non uniform character of the seeding pulse transverse profile is apparent in the far field distribution. In addition, the formation of multiple filaments can exacerbate the transverse non-uniformity of the plasma amplifier. We therefore speculate that the plasma defocusing effect also plays an important role for the saturation of the lasing yield when $P_{in}$ becomes larger than 1.5 TW.

To conclude, the above experimental observations show that upon increase of pump pulse energy, the density of population inversion keeps unchanged or increases, while the plasma amplifier volume increases considerably. Consequently, the gain saturation cannot be explained by complete depletion of the population inversion. To the contrary, our analysis indicates that there is room for an increase in lasing efficiency in air with TW lasers. On the other hand, the white-light seed pulse shows pronounced saturation and most probably experiences defocusing during its propagation in the amplifying plasma medium. We believe that these last two factors are at the origin of the lasing yield saturation.

In summary, we observed lasing action of atmospheric air pumped by 800 nm laser pulses with peak power up to 4 TW. The radiation yield first increases rapidly but saturates for pulse power larger than 1.5 TW. The maximum lasing radiation energy was 2.6 μJ and the maximum conversion efficiency was $3.5\times10^{-5}$. We attribute the lasing energy saturation to the sub-linear increase of the seed pulse with increasing pump power and the effect of plasma defocusing on the seed propagation.